%% file: template_lat22.tex
\documentclass[a4paper,11pt]{article}
\usepackage{booktabs}
\usepackage{pos}
\usepackage{subcaption}
\usepackage{graphicx}

\newcommand{\be}{\begin{equation}}
\newcommand{\ee}{\end{equation}}
\newcommand{\ba}{\begin{align}}
\newcommand{\ea}{\end{align}}
\newcommand{\baa}{\begin{array}}
\newcommand{\eaa}{\end{array}}
\newcommand{\bea}{\begin{eqnarray}}
\newcommand{\eea}{\end{eqnarray}}

\newcommand{\tl}{\tilde{l}}

\title{SU(N) fractional Instantons}
\ShortTitle{SU(N) fractional Instantons}

\author*[ab]{Jorge Luis Dasilva Gol\'an}
\author[a]{Margarita García Pérez}

\affiliation[a]{Instituto de F\'{\i}sica Te\'orica UAM-CSIC, Nicol\' as Cabrera 13-15,
    Campus de Cantoblanco.\\
    E-28049, Madrid, Spain}
\affiliation[b]{Departamento de F\'{\i}sica Te\'orica, m\'odulo 15, Universidad Aut\'onoma de Madrid, Cantoblanco.\\
    E-28049, Madrid, Spain}

\emailAdd{jorge.dasilva@uam.es}
\emailAdd{margarita.garcia@csic.es}

\abstract{\input{abstract.tex}}

\FullConference{%
The 39th International Symposium on Lattice Field Theory,\\
8th-13th August, 2022,\\
Rheinische Friedrich-Wilhelms-Universität Bonn, Bonn, Germany
}

\begin{document}
\begin{flushright}
  IFT-UAM/CSIC-22-147
\end{flushright}
\maketitle
	
\section{Introduction}
\input{intro.tex}

\section{Action density}
\label{sec:density}
\input{action.tex}

\section{Polyakov loops}
\label{sec:pol}
\input{pol.tex}

\section{Summary}
\label{sec:con}
\input{conclusions.tex}

\section*{Acknowledgments}
\addcontentsline{toc}{section}{Acknowledgments}

This work is partially supported by grant PGC2018-094857-B-I00 funded by “ERDF A way of making Europe”, by MCIN/AEI/10.13039/501100011033, and by the Spanish Research Agency (Agencia Estatal de Investigación) through grants IFT Centro de Excelencia Severo Ochoa  SEV-2016-0597 and  No CEX2020-001007-S, funded by MCIN/AEI/10.13039/501100011033. We acknowledge support from the project H2020-MSCAITN-2018-813942 (EuroPLEx) and the EU Horizon 2020 research and innovation programme,
STRONG-2020 project, under grant agreement No 824093. 
JDG acknowledges support under grant PRE2018-084489 funded by MCIN/AEI/ 10.13039/501100011033 and, by “ESF Investing in your future”.
We also acknowledge the use of the Hydra cluster at IFT and HPC resources at CESGA (Supercomputing Centre of Galicia).

\end{document}

%% file: abstract.tex
We present our study of a set of solutions to the $SU(N)$ Yang-Mills equations of motion with fractional topological charge. The configurations are obtained numerically by minimizing the action with gradient flow techniques on a torus of size $l^2\times(Nl)^2$ with twisted boundary conditions. We pay special attention to the large $N$ limit, which is taken along a very peculiar sequence, with the number of colors $N$ and the magnetic flux $m$ selected respectively as the $n$-th and $n-2$ terms of the Fibonacci sequence. We discuss the large $N$ scaling of the solutions and analyze several gauge invariant quantities as the Polyakov loops. We also discuss the so-called Hamiltonian limit, with one of the large directions sent to infinity, where these instantons represent tunneling events between inequivalent pure gauge configurations. 

%% file: intro.tex
The search for solutions to the Yang-Mills (YM) equations of motion on the hypertorus has had a long history starting with the seminal works of 't Hooft in the early 80's~\cite{tHooft:1979rtg}.
The freedom to select boundary conditions, corresponding to  gauge potentials periodic up to gauge transformations, leads to the 
appearance of new topological classes, implying, in particular, the possibility  of having fractional topological charge. 
In this paper, we will consider so-called twisted boundary conditions (TBC) whereby, under the shift by a torus period, the gauge potential transforms as:
\be
A_\mu(x + l_\nu \hat e_\nu) = \Omega_\nu(x) A_\mu(x)  \Omega_\nu^\dagger(x) + i \, \Omega_\nu(x) \partial_\mu \Omega_\nu^\dagger(x),
\label{eq:twisted_ec}
\ee
with $\Omega_\nu(x)$ taken as $SU(N)$ matrices constrained by the consistency relation:
\be
\Omega_\nu (x+l_\mu \hat e_\mu) \, \Omega_\mu(x) = \exp \{i 2 \pi n_{\nu \mu } /N\}\,  \Omega_\mu (x+ l_\nu \hat e_\nu)\,  \Omega_\mu (x),
\label{eq:consistency_x}
\ee 
and $n_{\mu \nu}$ a non-trivial antisymmetric tensor of integers defined modulo $N$, known as twist tensor. 

One interesting property that TBC introduces is that the topological charge $Q$ is no longer bounded to be an integer, but it is quantized in terms of the twist tensor as \cite{tHooft:1981nnx,vanBaal:1982ag}:
\be
Q= \frac{1}{16\pi^2} \int d^4 x \text{Tr} \left( F_{\mu \nu}(x)\widetilde F_{\mu \nu} (x)\right) = \nu - \frac {\vec k \cdot \vec m }{N},
\label{eq:qtopo}
\ee
where $ m_i =\epsilon_{ijk} n_{jk} /2$ is the spatial part of the twist, the so-called magnetic flux, while $k_i = n_{0i}$ is dual to the electric flux characterizing the Hilbert space in a Hamiltonian set-up.

Following these considerations, we summarize in this work our search for $SU(N)$ instanton solutions with fractional topological charge $Q=1/N$~\cite{DasilvaGolan:2022jlm}. We are particularly interested in those corresponding to a Hamiltonian set up: $\mathbf{R} \times \mathbf T^3$, with $\mathbf T^3$ a 3-dimensional torus endowed with TBC. 
In this limit, the fractional instantons we will present have a  well defined interpretation as tunneling events interpolating between two different pure gauge configurations at $x_0\rightarrow\pm\infty$. 
Fractional charge solutions of this type were first obtained numerically for $SU(2)$~\cite{GarciaPerez:1989gt,GarciaPerez:1997fq} and later generalized to $SU(N)$~\cite{GarciaPerez:1997fq,Montero:2000mv} for various choices of the twist tensor (vortex-like solutions have also been obtained~\cite{Gonzalez-Arroyo:1998hjb,Montero:2000pb}). In this proceedings, we focus on new $SU(N)$ solutions that have been obtained on asymmetrical tori with sizes $l_1=l_2=l/N$, $l_3=l$, and $l_0=sl$, with s a free parameter that is taken to infinity in the Hamiltonian limit. The number of colors is taken along the Fibonacci sequence as $N=F_{n}$, and the electric and magnetic fluxes are given respectively by: $\vec k = (0,0,-\bar m)$, and $\vec m= (0,0,m)$, with $m=F_{n-2}$ and $\bar m = (-1)^n F_{n-2}$; it is easy to check, using eq.~\eqref{eq:qtopo}, that this choice leads to topological charge $Q=1/N$, for $\nu = (-1)^{n+1} F_{n-4}$.

Let us explain the reasons behind these very particular choices.
The logic behind our geometry follows the idea of volume independence under TBC, see i.e.~\cite{Gonzalez-Arroyo:1982hyq,GarciaPerez:2014cmv,GarciaPerez:2020gnf} and references therein. With our choice of twist, the color and spatial degrees of freedom get entangled, and the torus periods in the twisted directions become effectively enlarged by a factor of $N$. Hence, the effective dynamics of our spatially asymmetric torus corresponds, in the large N limit, to a symmetric one with period $l$ in all three spatial directions.   
On the other hand, our selection of twist aims to avoid the appearance of tachyonic instabilities and center symmetry breaking in the large N limit~\cite{Ishikawa,Bietenholz:2006cz,Teper:2006sp,Azeyanagi:2007su,Guralnik:2002ru}. It has been conjectured that in order to avoid these problems the flux $m$ should be scaled with $N$~\cite{Gonzalez-Arroyo:2010omx} and this is optimally achieved by taking $m$ and the number of colors as the $n-2$ and $n$-th terms of the Fibonacci sequence~\cite{Chamizo:2016msz,GarciaPerez:2018fkj}. 

We have analyzed several gauge groups: $N=3,5,8,13,21$. Using gradient flow minimization,
numerical solutions are obtained on lattices of size: $s NL  \times L^2 \times NL  $. Fixing the lattice spacing through: $a = l/N L$, our discretization corresponds to a continuum torus with periods $s l  \times (l/N)^2 \times l  $ with TBC implemented in the usual way.

\begin{figure}[t]
\centering
\includegraphics[width=0.5\linewidth]{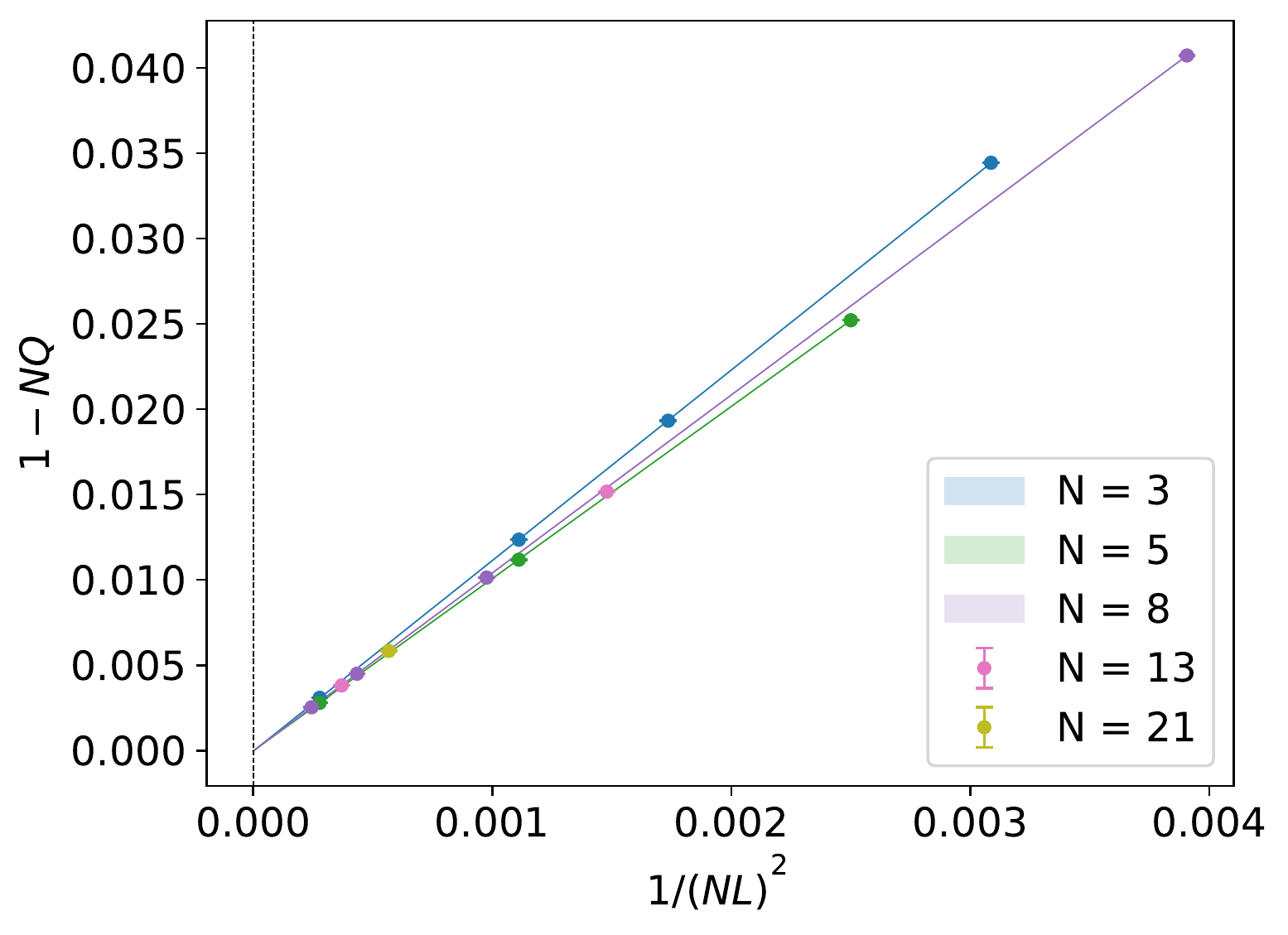}
\caption{Continuum extrapolation of $1-NQ$ with $a^2= 1/(NL)^2$ corrections.}
\label{fig:extrap}
\end{figure}

For each value of N (and for different lattice spacings) we have computed the total action, the electric and magnetic parts of the action (computed to test self-duality), and the topological charge. Selecting  different values of the lattice spacing, we were able to extrapolate these quantities to the continuum. As an illustration, figure~\ref{fig:extrap} shows the extrapolation of $1-NQ$ that agrees with the continuum result up to one part in $10^4$.

In what follows, section~\ref{sec:density} discusses the action density of the solutions, comparing them with the constant curvature fractional instantons that are know to exist for certain values of the torus aspect ratios~\cite{tHooft:1981nnx,Gonzalez-Arroyo:2019wpu};
section~\ref{sec:pol} analyzes Polyakov loops and we conclude with a brief summary of results.

%% file: action.tex
\begin{figure}[t]
\centering
\includegraphics[width=0.9\linewidth]{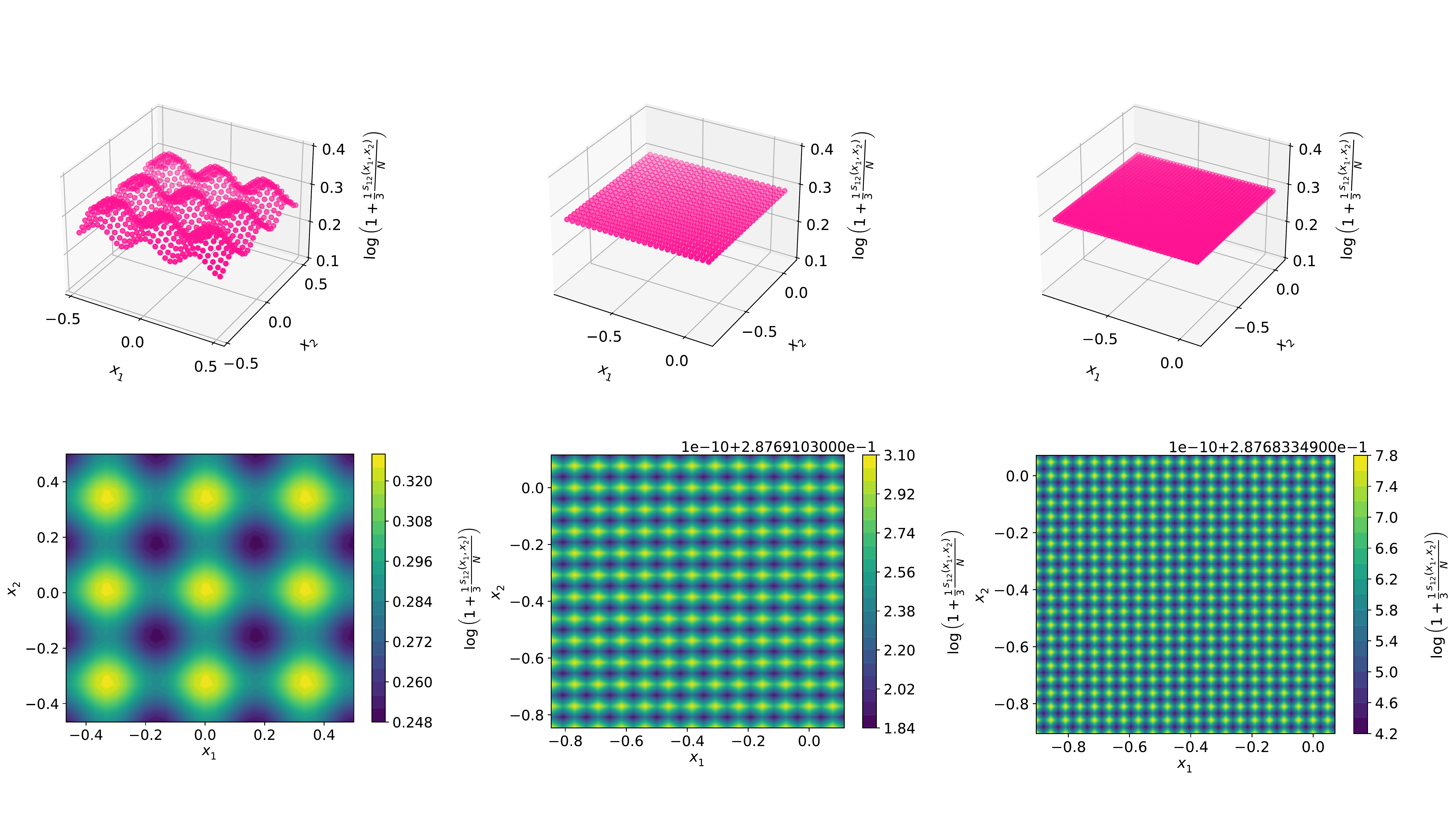}
\caption{Profile $l^2s_{12} (x_1,x_2)/N$ as a function of $x_1/l$ and $x_2/l$. In order to make the structure more visible, the quantity displayed is $\log(1+l^2 s_{12}(x_1,x_2)/(3N)$. Gauge groups are, from left to right: $SU(3)$, $SU(13)$, and $SU(21)$.}
\label{fig:q12}
\end{figure}
\begin{figure}[h]
\centering
\includegraphics[width=0.9\linewidth]{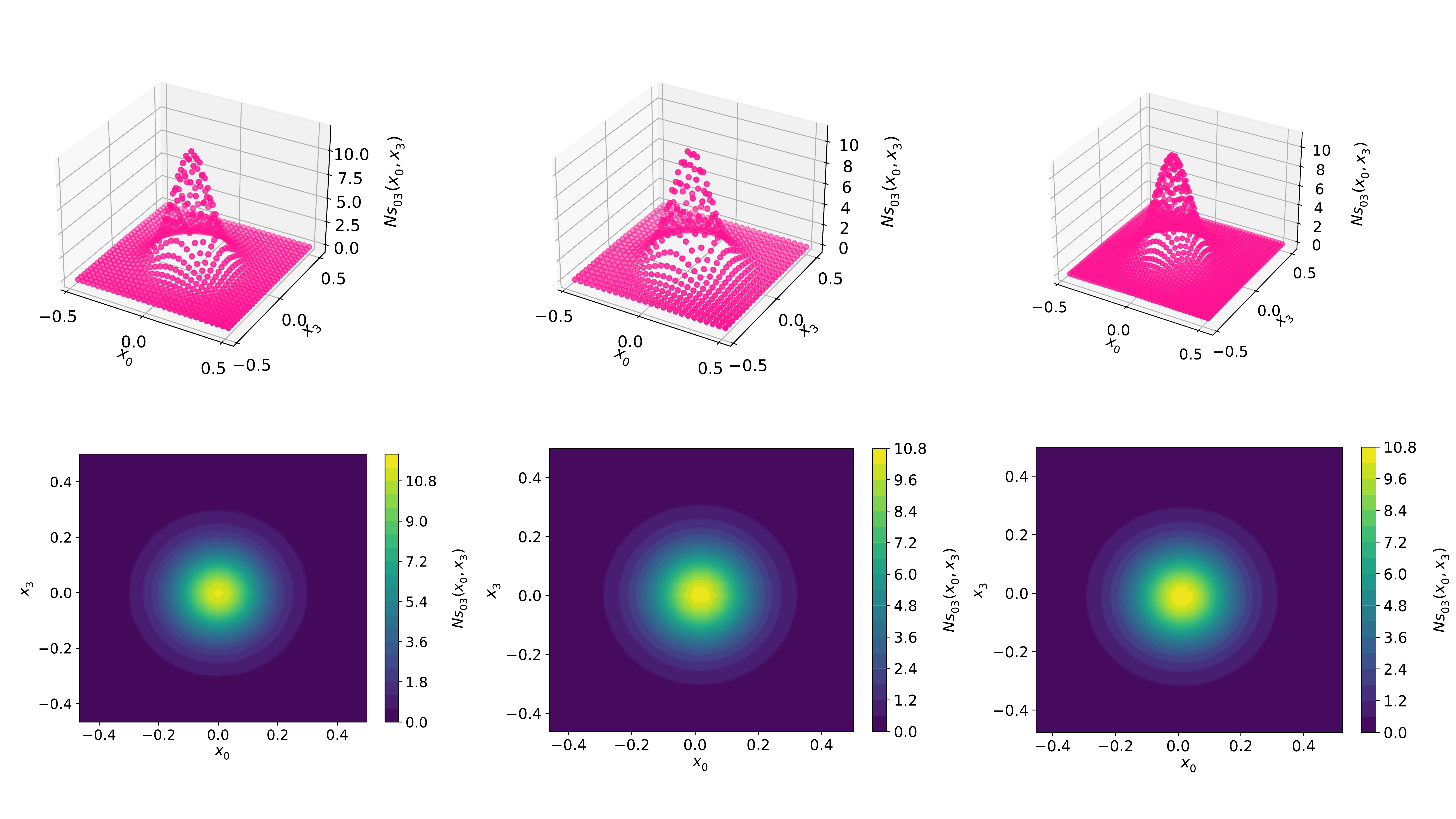}
\caption{Profile $N l^2s_{03} (x_0,x_3)$ as a function of $x_0/l$ and $x_3/l$. Gauge groups are, from left to right: $SU(3)$, $SU(13)$, and $SU(21)$.
We have taken $l=1$ to set the scale. }
\label{fig:s03-l0}
\end{figure}

We have determined the density profiles (in units of $8\pi^2$) obtained by integrating the 4-dimensional action density
along two or three spatial directions:
\begin{align} 
 s_{\mu \nu} (x_\mu,x_\nu) &\equiv \left (\prod_{\rho \ne \mu, \nu}\int_0^{l_\rho} d x_\rho \right ) \,  s(x),
\label{eq:profiles-2d}\\
 s_\mu(x_\mu)  &\equiv \left (\prod_{\rho \ne\mu} \int_0^{l_\rho} d x_\rho\right ) \, s(x).
\label{eq:profiles-1d}
\end{align}
In the analysis of the large $N$ scaling of these profiles, an inspiring guide has been the case of constant curvature fractional charge solutions that are known to exist for certain values of the torus periods. Following the general construction presented in ref.~\cite{Gonzalez-Arroyo:2019wpu}, it is easy to obtain solutions of this type with gauge group and twist parameters taken in the Fibonacci sequence  
-- a complete deduction can be found in ref. \cite{DasilvaGolan:2022jlm}. The important result is that taking $m=F_{n-2}$ and $N=F_{n}$, solutions to the Yang-Mills self-duality equations exist for any value of $l_0$ satisfying:
\be
l_0/l = \frac{F_{n-m+1} F_{n-m}} {F_n^2 F_{m} F_{m-1}}
\label{eq:fibonacci-cc}
\ee
where we have scaled $l_1=l_2=l/F_n$, and set $l_3=l$. For these solutions the profiles are flat and satisfy: $l^2s_{12}(x_1,x_2)/N=1$; 
$l^2 N s_{03}(x_0,x_3)= l/l_0$; and $l N s_0(x_0) = l/l_0$. In the limit $n\rightarrow \infty$, the maximal value of the time extent is attained for $l_0/l=\varphi^{-3} \sim 0.236$, with $\varphi$ the Golden Ratio. 
We will see below how our solutions compare with these estimates.

Let us start with the analysis of 2-dimensional profiles. Fig.~\ref{fig:q12} shows $l^2 s_{12}(x_1,x_2)/N$ as a function of the coordinates in the twisted plane for several values of $N$; each configuration has been replicated $N$ times in each direction for the display. 
Profiles are practically flat in this plane, with the level of flatness increased as the value of $N$ grows, becoming almost independent of the two short directions.
In addition, the hight of the profiles matches very well the one expected for the constant curvature solution which for the quantity displayed in the plot is $\log(1+1/3) = 0.287682$.

Figure~\ref{fig:s03-l0} shows the profiles of $Nl^2s_{03} (x_0,x_3)$, for $l_0=l_3=l$ ($s=1$). In this case, the solutions are localized, developing a maximum and decaying fast far away from the center. Notice that the hight of the peak is not far from the value expected for the constant curvature solution with maximal value of $l_0/l$, i.e. $l/l_0 \sim 4.427$; we will come back to this point later on.

\begin{figure}[t]
\begin{subfigure}{.45\textwidth}
  \centering
  \includegraphics[width=\linewidth]{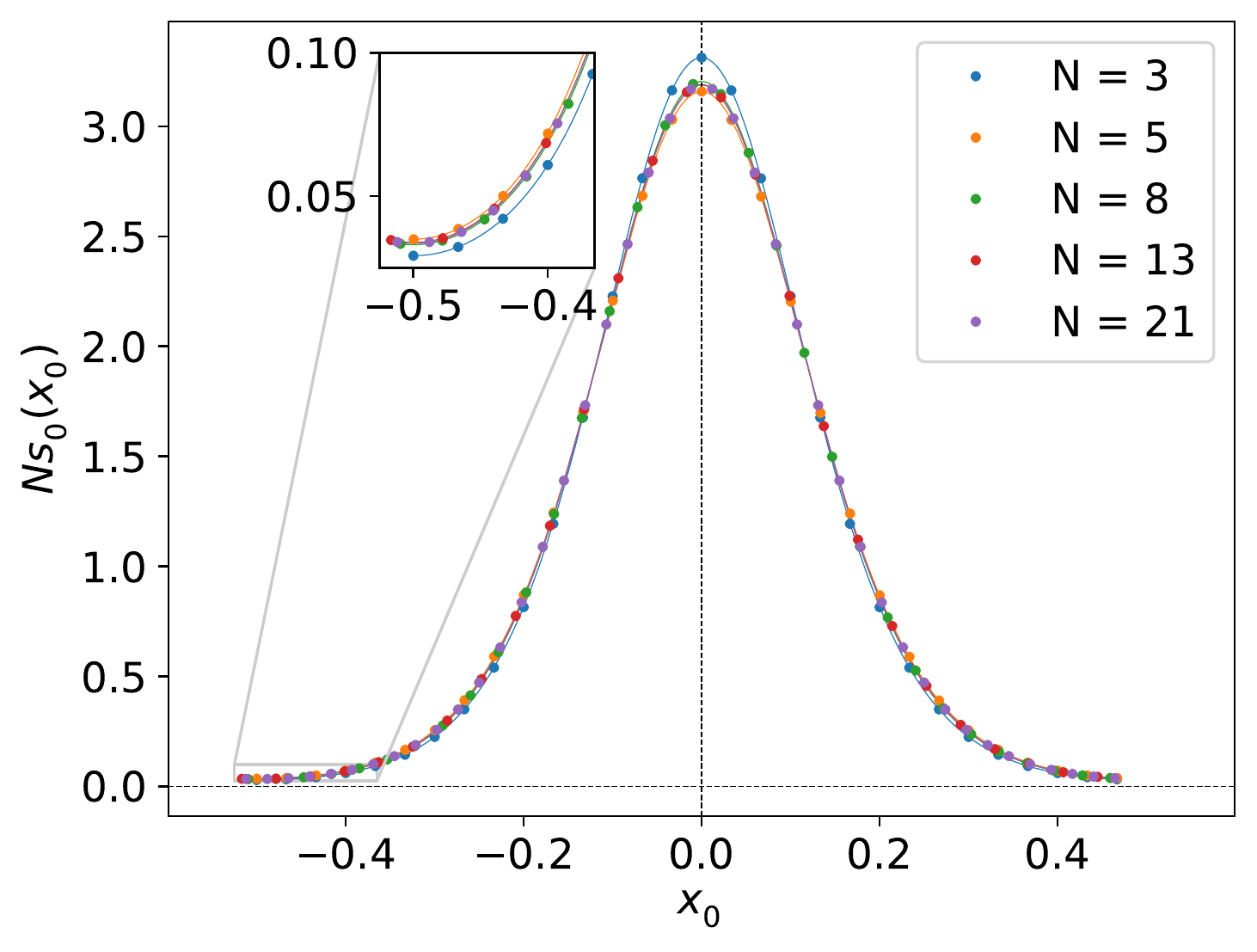}
  \caption{$s=1$}
  \label{fig:q0-variousN}
\end{subfigure}%
\begin{subfigure}{.45\textwidth}
  \centering
  \includegraphics[width=\linewidth]{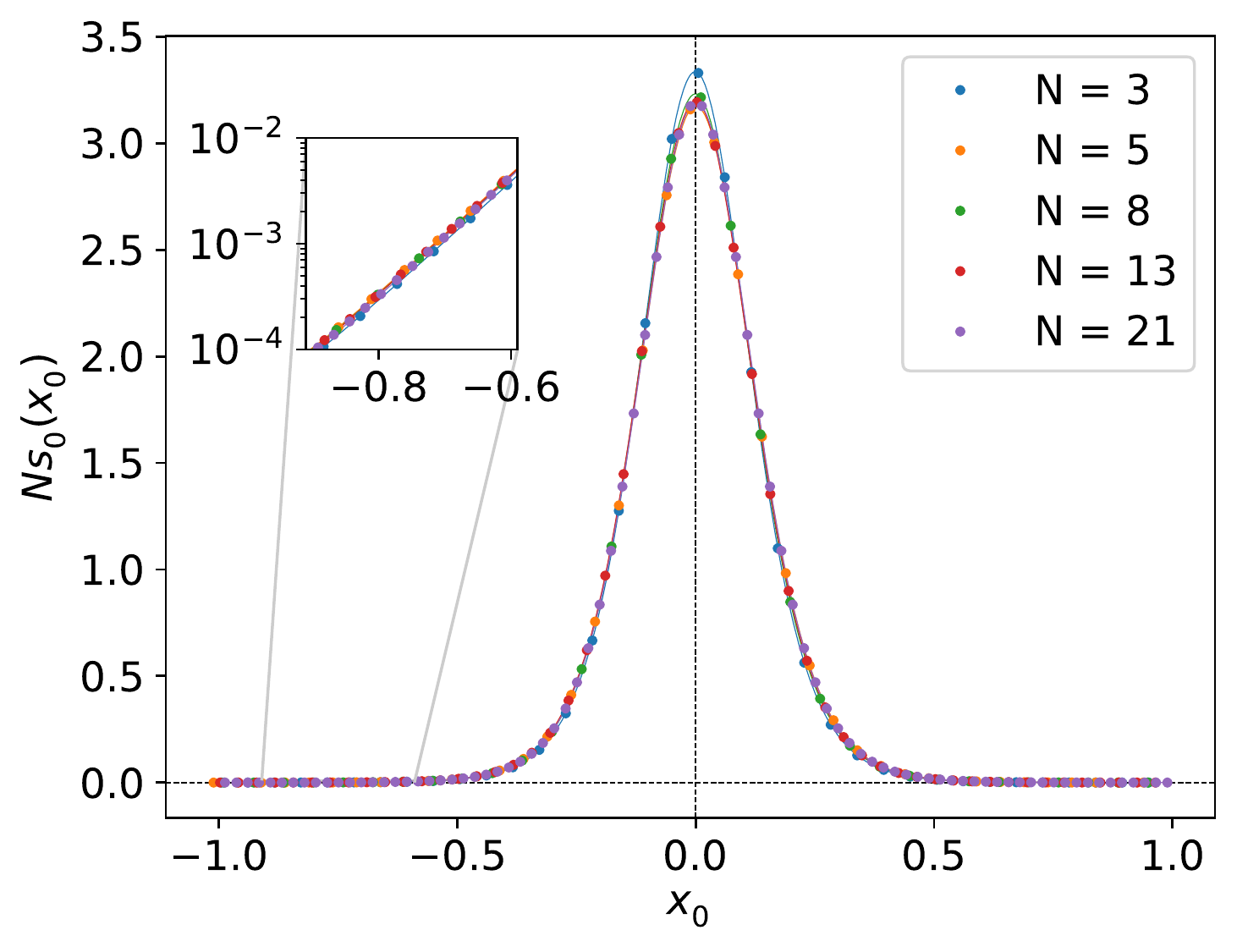}
  \caption{$s=2$}
  \label{fig:q0-variousN-2l0}
\end{subfigure}
\caption{Time dependence of the energy density profile for various gauge groups $N$ and time extents $l_0=sl$, with $s=1$ and 2. The inset in the right plot is in logarithmic ($y-$) scale to show that the decay in the tails is exponential in time.}
\end{figure}

We can obtain a more quantitative comparison if we look at the 1-dimensional energy profiles, obtained by integrating over the 3 spatial coordinates. Figs.~\ref{fig:q0-variousN} and~\ref{fig:q0-variousN-2l0} show the value of $N l s_0(x_0)$ as a function of $x_0/l$ for  various values of $N$ and $s=1$ and $s=2$ respectively. It is clear that, once the overall factor $N$ is set, the remaining $N$ dependence is rather small, and all curves tend to a universal behavior. Moreover, enlarging the time direction with a factor $s=2$ does not change drastically the shape of the profile, except at the tails where it decays exponentially. 

\begin{figure}[t]
  \centering
  \includegraphics[width=0.7\linewidth]{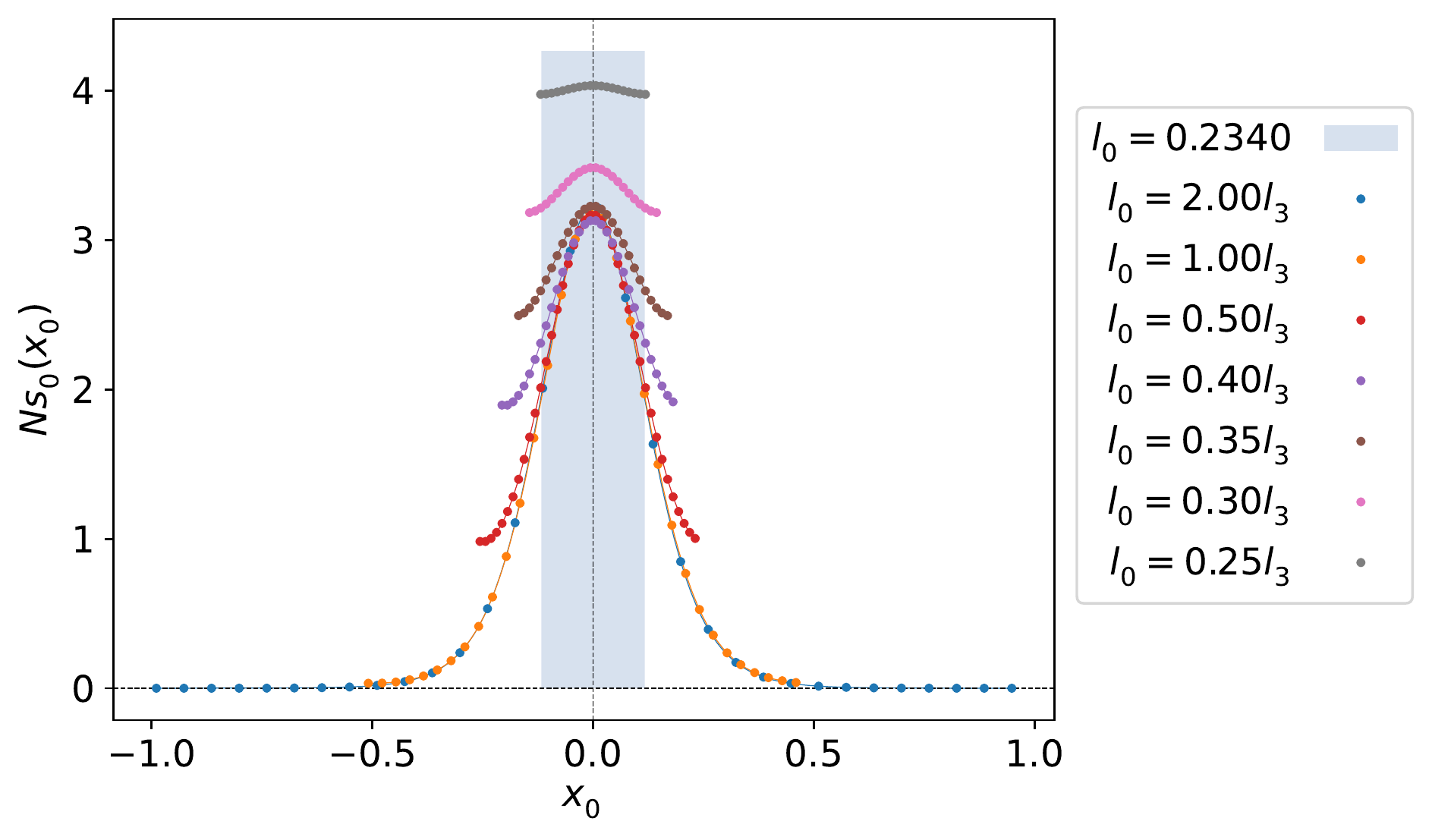}
\caption{Several solutions for different values of time extent $l_0$ compared to the constant curvature one with $l_0/l=\varphi^{-3}=0.236$. }
\label{fig:flat}
\end{figure}

To compare our solution with the constant curvature one, we have generated several minimum-action configurations starting at $l_0=0.25l$ (close to the Fibonacci case), and growing up to $l_0=l$. The resulting energy density profiles are shown in figure~\ref{fig:flat}. The solutions change rapidly at the beginning, but beyond $l_0\sim0.5 l$ they remain unchanged except at the tails where, in the large $l_0/l$ limit, they  decay exponentially.
We believe this exercise is very clarifying, going in the line of interpreting these solutions as deformations of the constant curvature ones~\cite{Gonzalez-Arroyo:2019wpu,GarciaPerez:2000aiw}.

\begin{figure}[t]
\begin{subfigure}{.49\textwidth}
  \centering
  \includegraphics[width=\linewidth]{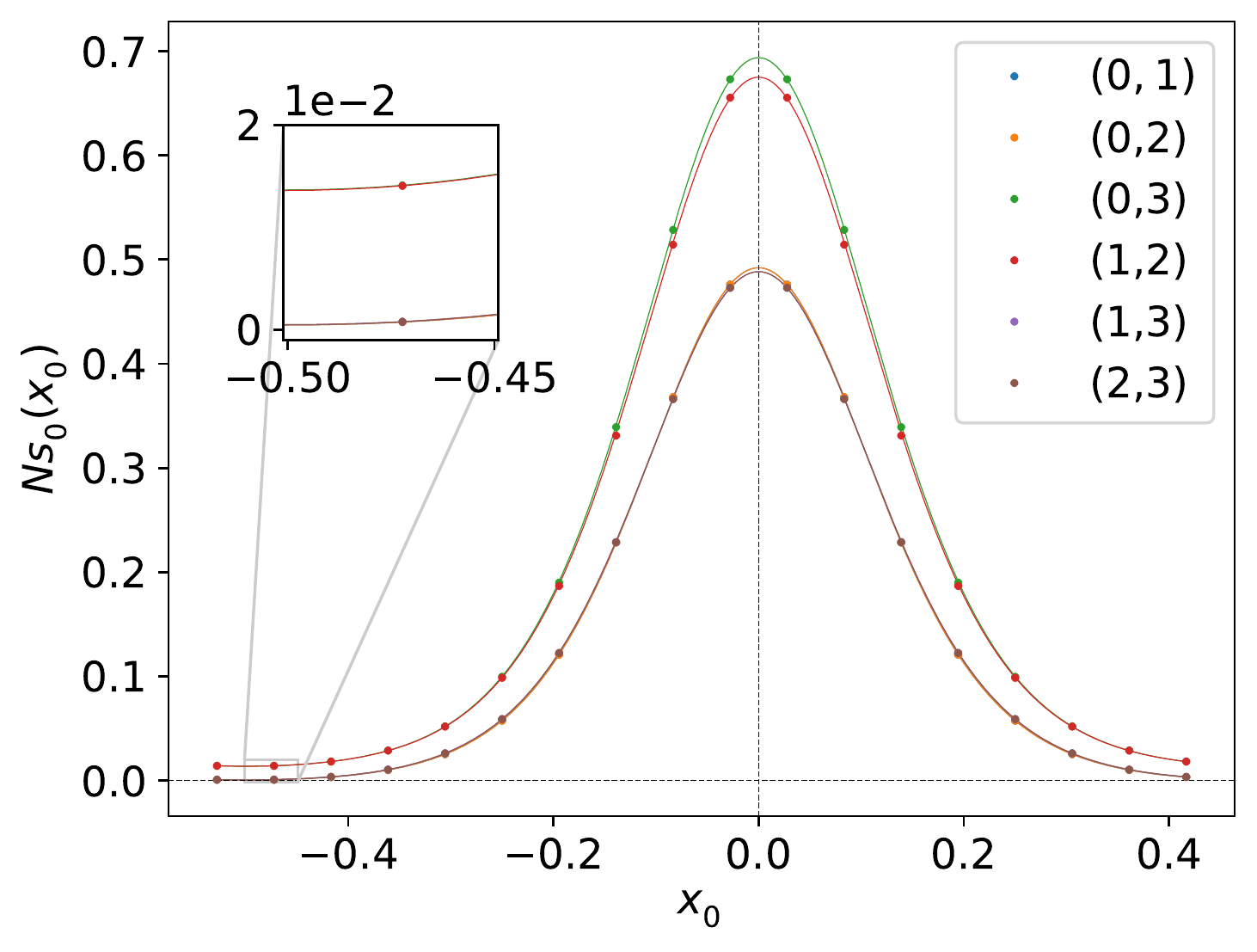}
  \caption{SU(3) with $L=6$ and $s=1$}
\end{subfigure}
\begin{subfigure}{.49\textwidth}
  \centering
  \includegraphics[width=\linewidth]{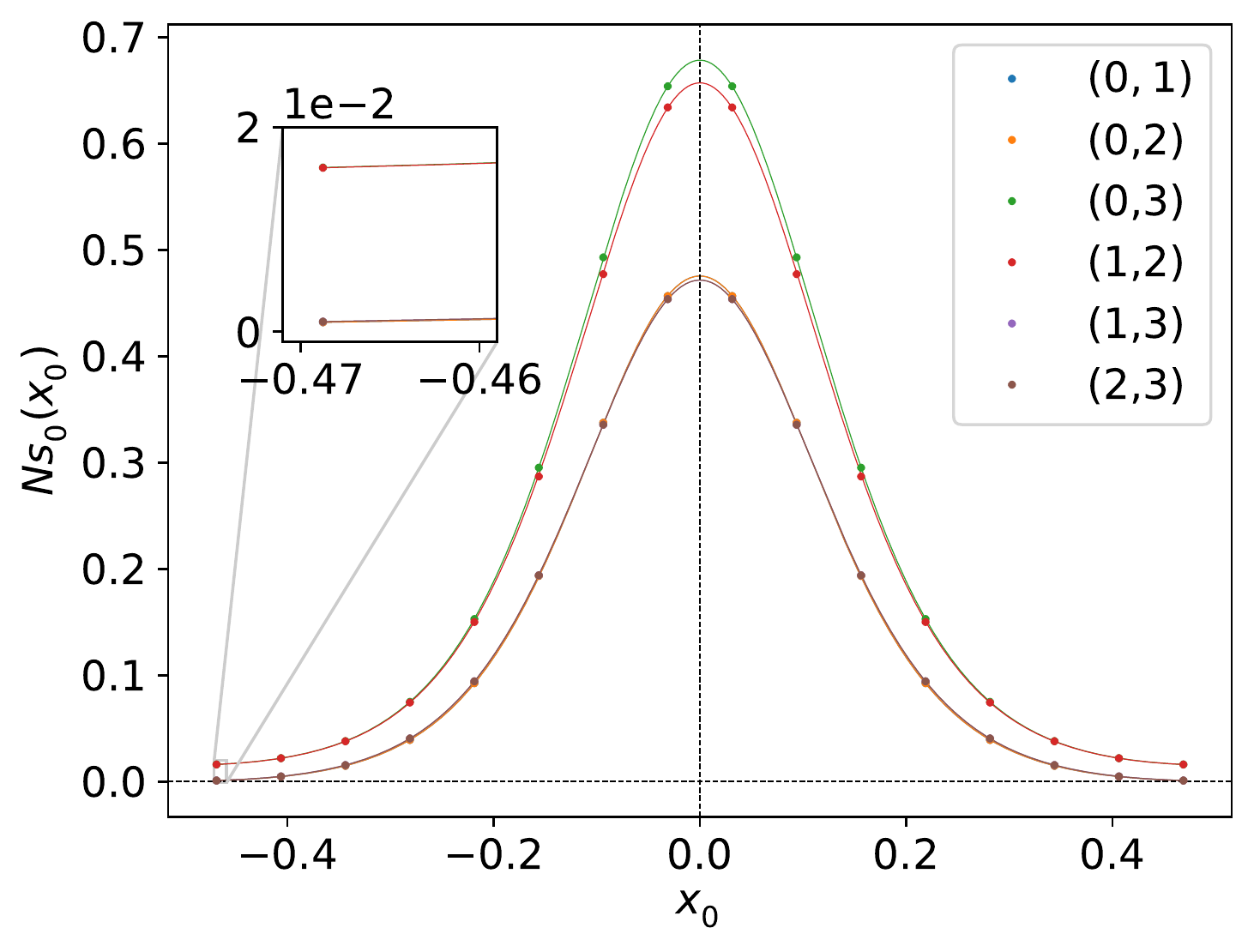}
  \caption{SU(8) with $L=2$ and $s=1$}
\end{subfigure}
\begin{subfigure}{.49\textwidth}
  \centering
  \includegraphics[width=\linewidth]{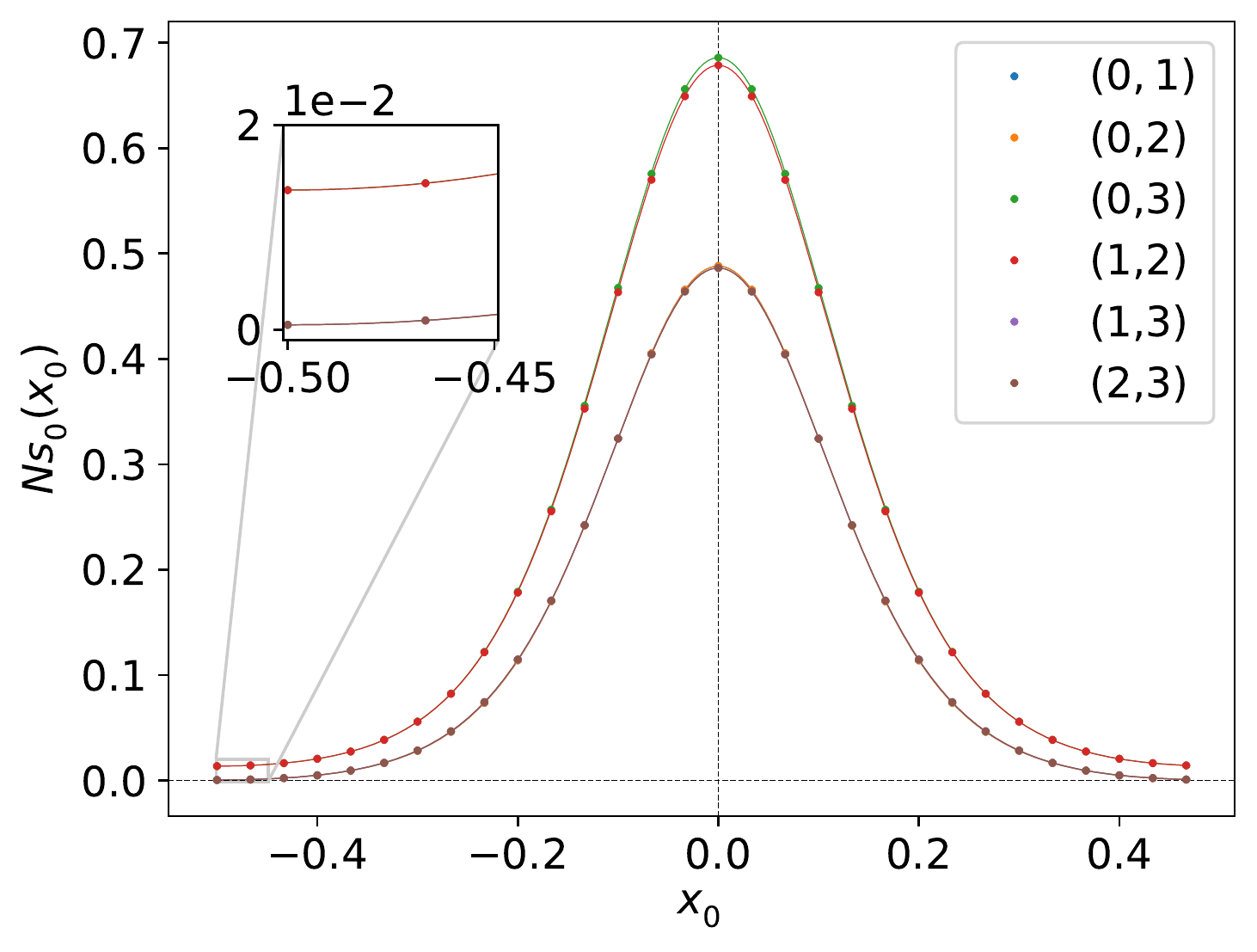}
  \caption{SU(3) with $L=10$ and $s=1$}
\end{subfigure}
\begin{subfigure}{.49\textwidth}
  \centering
  \includegraphics[width=\linewidth]{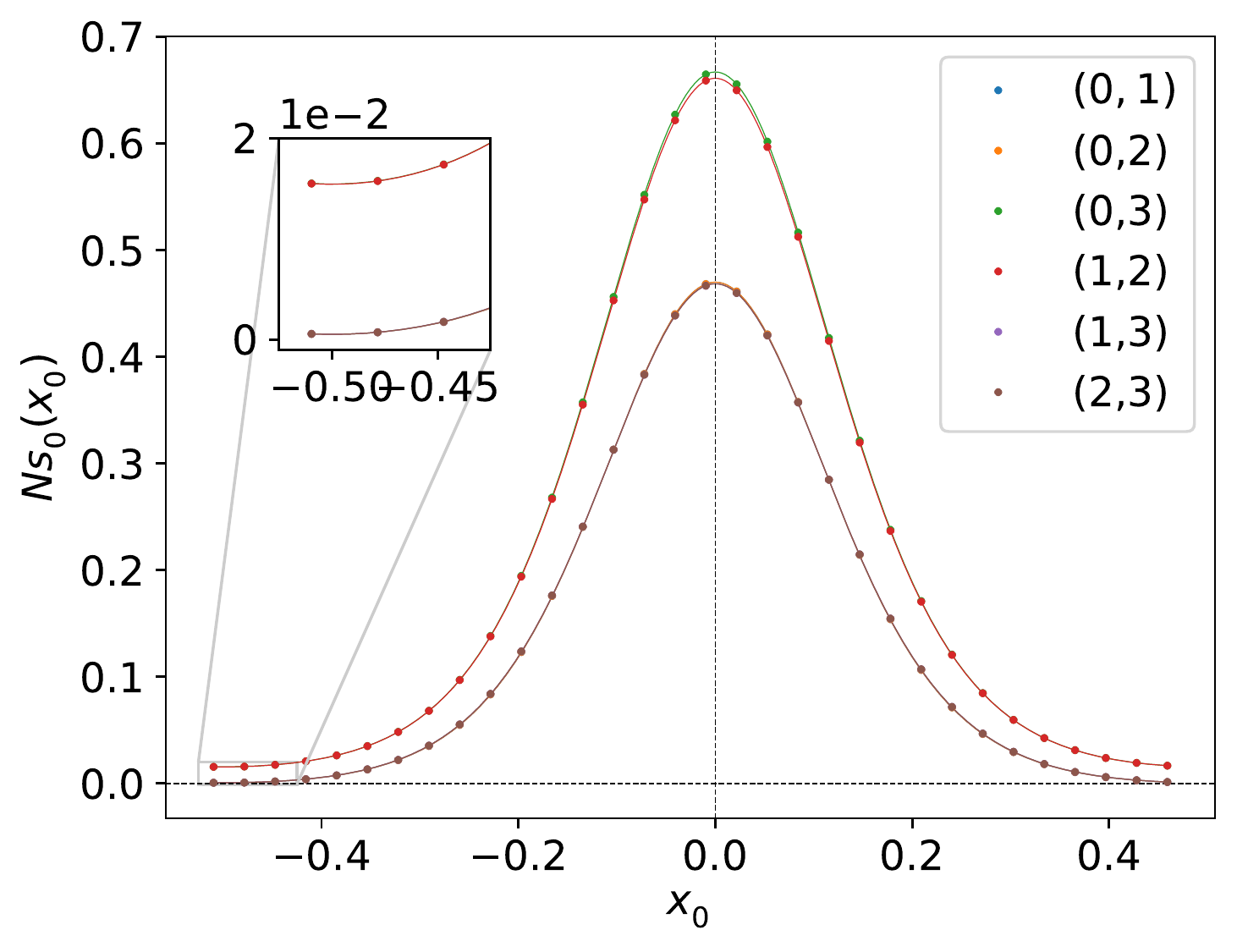}
  \caption{SU(8) with $L=4$ and $s=1$}
\end{subfigure}
\caption{Electric and magnetic components of the energy density as a function of $x_0$, obtained by integrating $\text{Tr}(F_{\mu \nu}^2)$ for different values of $N$ and lattice spacings.}
\label{fig:self-dual}
\end{figure}

One final test that can be done concerns self-duality. To test this, we have computed the energy profiles in one direction, but separating the computation of the electric and magnetic contributions as $\rm{Tr}(F_{\mu \nu}^2)$, with $\mu$ and $\nu$ fixed. The result for several values of $N$ and various lattice spacings is shown in fig.~\ref{fig:self-dual}. The degree of self-duality is very high in all configurations and tends to increase as the continuum limit $a=l/(LN) \rightarrow 0$ is approached. 

%% file: pol.tex
In this section we will analyze the Polyakov loops, and, based on the Hamiltonian limit discussed in the introduction, we will check the  interpretation of the fractional instantons as tunneling events interpolating between two pure gauge configurations. 

Let us start the discussion by defining $P_\mu(x)$ as ($1/N$ times) the trace of the Polyakov loop winding the torus once in direction $\mu$:
\be
P_\mu(x) = \frac{1}{N} \rm{Tr} \Big( P\exp \Big\{ -i \int_0^{l_\mu} dx_\mu A_\mu(x)\Big \}\, 
\Omega_\mu(x)\Big ) \equiv |P_\mu(x) | \, e^{i\phi_\mu(x)}.
\ee
Notice that this definition is slightly different from the case with periodic boundary conditions, needing the insertion of a $\Omega_\mu(x)$ matrix to preserve gauge invariance. The periodicity properties of the loop derived from this definition amount to:
\be
P_\mu (x+ l_\nu \hat e_\nu) = e^{i\frac{2 \pi n_{\mu \nu}}{N}} P_\mu (x).
\ee  

Once the twist is fixed, we can use gauge invariance to bring the three spatial $\Omega_i(x)$ matrices to a constant form,
the so-called twist eaters, that satisfy the consistency relation:
\begin{align}
\Gamma_1 \Gamma_2 &= e^{i \frac{2 \pi m}{N}} \Gamma_2 \Gamma_1,
\label{eq:twist1}\\
\Gamma_3 \Gamma_i &= \Gamma_i \Gamma_3 \text{, for }  i=1,2.
\label{eq:twist3}
\end{align}
It is trivial to check that $A_i=0$ is consistent with these boundary conditions and leads to Polyakov loops given by:
\be
P(\gamma, w_1, w_2,w_3) = \frac{1}{N} \rm{Tr} \left (\Gamma_1^{w_1(\gamma)}  \Gamma_2^{w_2(\gamma)}  \Gamma_3^{w_3(\gamma)} \right),
\ee
where $\gamma$ is a closed curve and $w_i(\gamma)$ is the winding in the $ith$ direction, defined modulo $N$. In our particular case, this implies $P(\gamma, 0,0,1) = z_3$ and $P(\gamma, w_1,w_2,0)=0$, unless $w_1$ and $w_2$ are both equal to zero (mod $N$), meaning that flat connections can be characterized with the value of the non-trivial Polyakov loop in the third direction.

\begin{figure}[t]
  \centering
  \includegraphics[width=0.60\linewidth]{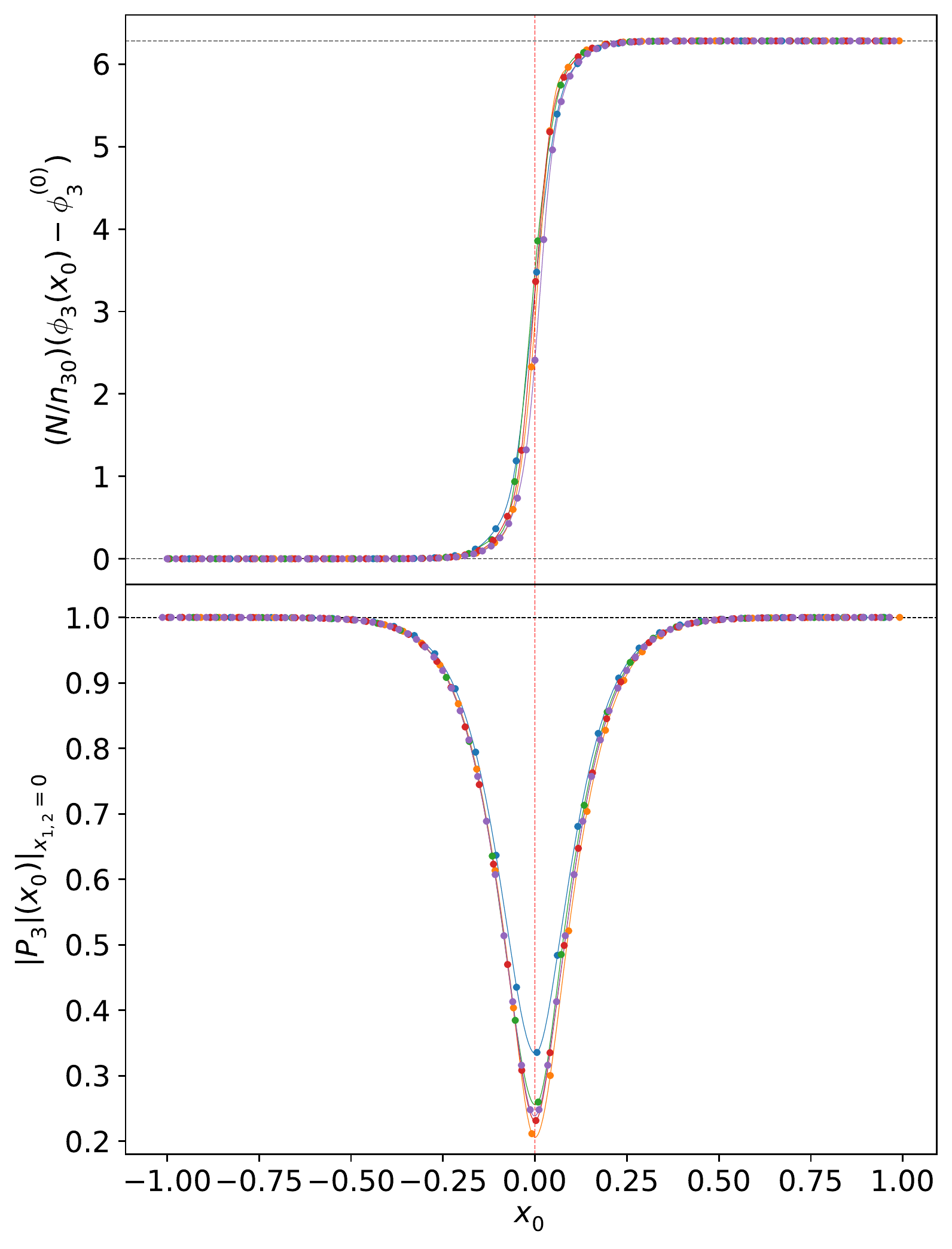}
\caption{We display as a function of time, for $l_0=2l_3=2l$,  and from top to bottom: the change of phase of $P_3$ multiplied by $N/n_{30}$ and its modulus, both evaluated at $x_1=x_2=0$.}
\label{fig:P3}
\end{figure}

Starting from a pure gauge configuration at $x_0=-\infty$, our choice of space-time twist combined with the periodicity condition of the Polyakov loop, leads to another pure gauge configuration at $x_0=+\infty$ that differs from the former 
in the value of the holonomy in the $x_3$ direction by a a factor $\exp\{2 \pi i n_{30}/N \}$. 
In what follows we will compare these expectations to our numerical results.

The two holonomies in the short torus directions remain all the time very small and tend to zero as we go in the time direction far from the instanton center (details are given in ref.~\cite{DasilvaGolan:2022jlm}).  At all values of $t$, the modulus of the two loops tends to decrease as the value of $N$ grows.

On the other hand, fig.~\ref{fig:P3} displays from top to bottom: the change of phase in $P_3$ (multiplied by $N/n_{30}$) and the modulus $|P_3|$, all of them for tori with $l_0=2l_3$. The results converge to  universal curves for all values of $N$ that match well the expectations. First, far from the instanton center, $P_3$ approaches an element of the center group, with modulus 1 and phase equal to $2 \pi /N$ times an integer. Moreover, the jump from $x_0=-\infty$ to $x_0=+\infty$ is as expected for our choice of space-time twist, with phases at both ends differing by a factor $2 \pi n_{30} /N$.

%% file: conclusions.tex
In this work, we have obtained a new type of $SU(N)$ instanton configuration on the hypertorus with the special property of having fractional topological charge $Q=1/N$, which is a direct consequence of the choice of twisted boundary conditions. We have analyzed several quantities, such as the action-density profile, or some gauge invariant observables such as the Polyakov loop. After an appropriate reescaling taking into account the leading $N$ dependence, the solutions show a universal pattern with a small remnant dependence on $N$ that goes away in the large $N$ limit. We have discussed how these new solutions relate to the constant curvature ones, that exist only for particlular values of the torus aspect ratios, suggesting that they could be obtained as smooth deformations of those along the lines presented in refs.~\cite{Gonzalez-Arroyo:2019wpu,GarciaPerez:2000aiw}. Finally, let us mention that we have payed particular attention to the Hamiltonian limit: $\mathbf R \times \mathbf T^3$, where these solutions represent tunneling events between two pure gauge configurations, 
essential to push the analytical perturbative domain beyond the small volume regime -- see i.e.~\cite{GarciaPerez:1993jw,RTN:1993ilw,GarciaPerez:2009mmu,vanBaal:2000zc} and references therein.